\documentclass[5p,sort&compress]{elsarticle}

\usepackage{graphicx}
\usepackage{subcaption}
\usepackage{hyperref}
\usepackage{amsmath}
\usepackage{natbib}
\usepackage{pifont}
\usepackage{geometry}
\usepackage{fleqn}
\usepackage{txfonts}
\usepackage{xcolor}
\usepackage{lineno}

\begin{document}

%\linenumbers

\title{Low-energy neutron cross-talk between organic scintillator detectors}

\author[LPC]{M. S\'enoville \fnref{addSenoville}}
\fntext[addSenoville]{Present address: Université Paris-Saclay, INRAE, AgroParisTech, UMR SayFood, Palaiseau F-91120, France}
\author[LPC]{F. Delaunay\corref{corr}} \ead{delaunay@lpccaen.in2p3.fr}
\cortext[corr]{Corresponding author}
\author[LPC]{N. L. Achouri}
\author[LPC]{N. A. Orr}
\author[LPC]{B. Carniol}
\author[IJC]{N. de S\'er\'eville}
\author[LPC]{D. \'Etasse}
\author[LPC]{C. Fontbonne}
\author[LPC]{J.-M. Fontbonne}
\author[LPC]{J. Gibelin}
\author[LPC]{J. Hommet}
\author[DAM]{B. Laurent}
\author[DAM]{X. Ledoux \fnref{addLedoux}}
\fntext[addLedoux]{Present address: GANIL, CNRS/IN2P3, CEA, Caen, France}
\author[LPC]{F. M. Marqu\'es}
\author[CIEMAT]{T. Mart\'inez}
\author[LPC]{M. P\^arlog}
\address[LPC]{Université de Caen Normandie, ENSICAEN, CNRS/IN2P3, LPC Caen UMR6534, F-14000 Caen, France}
\address[DAM]{CEA, DAM, DIF, F-91297 Arpajon, France}
\address[IJC]{Universit\'e Paris-Saclay, CNRS/IN2P3, IJCLab, 91405 Orsay, France}
\address[CIEMAT]{CIEMAT, Madrid, Spain}

\begin{abstract}
A series of measurements have been performed with low-energy monoenergetic neutrons to characterise cross-talk between two organic scintillator detectors. Cross-talk time-of-flight spectra and probabilities were determined for neutron energies from 1.4 to 15.5 MeV and effective scattering angles ranging from $\sim$50$^\circ$ to $\sim$100$^\circ$.
Monte-Carlo simulations incorporating both the active and inactive materials making up the detectors showed reasonable agreement with the measurements. Whilst the time-of-flight spectra were very well reproduced, the cross-talk probabilities were only in approximate agreement with the measurements, with the most significant discrepancies ($\sim$40 \%) occurring at the lowest energies.
The neutron interaction processes producing cross-talk at the energies explored here are discussed in the light of these results.
\end{abstract}


\maketitle

%\onecolumn

\section{Introduction}

The advent and development in recent decades of dedicated radioactive beam facilities have allowed for the production and investigation of the properties of nuclei lying far from the line of $\beta$-stability.  In the case of neutron-rich systems, $\beta$-decay $Q$-values are often high enough to permit decay to feed neutron unbound states in the daughter nucleus, and close to the dripline, $\beta$-delayed multi-neutron emission becomes possible.

Beyond determining neutron emission probabilities, information may be obtained regarding the decay scheme and structural properties by measuring the energies of the emitted neutrons.  This is most commonly performed using time-of-flight (TOF) measurements, whereby the neutron is detected in an organic scintillator (with the TOF start provided by the $\beta$-decay electron).  For such measurements modular neutron arrays, composed of a series of relatively thin ($\sim$2-5 cm), long ($\sim$60-200 cm), large-area ($\sim$200-5000 cm$^2$) plastic scintillator bars \cite{Harkewicz,TONNERRE,Hirayama,VANDLE} or a set of thin scintillator modules \cite{EDEN,LEND,MONSTER}, are commonly employed.   Such detectors offer good granularity, reasonably high intrinsic detection efficiency and good timing resolution. In principle, the use of multi-detector arrays facilitates the detection of events comprising two or more neutrons.

Of critical importance in such measurements, however, are the effects of cross-talk, whereby a single neutron incident on one detector is scattered into a second detector and the signal produced in each is sufficient for an event to be registered in both of them, thus mimicking the direct detection of two neutrons.
Cross-talk has been reasonably well characterised for neutron energies of $\sim$10-70~MeV \cite{Desesquelles,Cronqvist,Edel,DEMON,Wang,Kohley}, primarily for the purposes of reaction studies. More recently, such studies have been extended to much higher energies on the order of $\sim$100-300~MeV \cite{NakamuraNIM2016,KondoNIM2020}.
As evidenced by reaction studies at medium and high energies,
a clear and reliable understanding of cross-talk at the energies of interest is crucial for planning measurements (e.g., geometrical configuration of the array to maximise detection efficiencies and reduce rates of cross-talk), for the analysis (e.g., cross-talk rejection procedures) and interpretation of the results (e.g., reliable simulations to include the residual effects of the cross-talk events that cannot be eliminated in the analysis) \cite{Wang,Pluta,GhettiCT,MarquesCT,Marques2n,NakamuraNIM2016,KondoNIM2020}.

At energies below $\sim$10~MeV, however, there is a clear paucity of data available to enable low-energy cross-talk to be properly characterised and reliably simulated.  The very few measurements that have been made have deduced cross-talk rates integrated over a wide range of neutron energies or between long scintillator bars located in a close-packed geometry \cite{Saneesh,TONNERRE}.  As such, detailed systematic measurements of cross-talk at low energies are clearly required, in particular to allow for multi-neutron detection to be reliably extended to $\beta$-delayed neutron emission \cite{Delaunay} and to aid other applications, such as the measurement of neutrons from fusion-evaporation reactions \cite{NEDA} or from spontaneous fission \cite{Joyce}.

Here we report on such dedicated measurements of cross-talk probabilities and TOF spectra between two liquid scintillator detector modules for a range of configurations $-$ scattering angles from $\sim$50$^\circ$ to $\sim$100$^\circ$ $-$ using monoenergetic neutrons in the range 1.4 to 15.5 MeV.  The results are compared with detailed Monte-Carlo simulations.  In the light of these, the scattering and reaction processes leading to cross-talk are discussed.

\section{Experiment}

The present measurements were performed at the 4 MV Van de Graaff facility of the CEA/DIF research centre (Arpajon, France).
Monoenergetic neutrons were produced by reactions induced by proton and deuteron beams on thin titanium deuteride (TiD$_x$) and titanium tritide (TiT$_x$) targets.
The energies of neutrons produced at 20$^\circ$ from the beam axis and used for cross-talk measurements are given in Table \ref{TabBeamsTargets}, together with details on the reactions, beams and targets employed.
The targets were produced by SODERN \cite{SODERN} by physical vapour deposition of titanium onto a 1.5-mm thick gold backing, followed by impregnation of the titanium layer by deuterium or tritium gas \cite{NIM_SODERN}.
The charged particle beams stopped in the target backing. The beam energies were below the Coulomb barrier energies for titanium and gold, therefore no nuclear reactions on the titanium film or backing could produce neutrons. The neutron energy spread was determined by the energy loss of the beam in the target. It was of a few tens of keV, except for the neutron energy 15.48 MeV where it reached a few hundreds of keV.
As described below, the charged particle beams were pulsed to enable TOF measurements to be performed.
Their intensities were adjusted to obtain neutron detection rates between a few hundred and a few thousand per second in the detector placed at 20$^\circ$.

The measurements were performed in two experiments, with a pair of identical neutron detectors.
In the first experiment, a pair of modules of the EDEN array \cite{EDEN} was used, whilst the second experiment employed MONSTER modules \cite{MONSTER}. Both detector designs are based on a cell with a diameter of 20 cm and a thickness of 5 cm, filled with liquid scintillator (NE213 for EDEN, and the equivalent BC501A for MONSTER), and a photomultiplier tube (PMT) with a diameter of 12.7 cm. The cell and the PMT are coupled together via a tapered polymethyl methacrylate (PMMA) light guide.
The two detector types differ primarily in the thickness of the light guide, which is 50 mm for the EDEN modules and 30 mm for the MONSTER detectors, and in details of the casing materials and thicknesses.
The light guide is fixed on the glass window closing the scintillator cell on one side, whilst the opposite side is closed by the thin, opaque entrance window. The latter is made of stainless steel in EDEN modules and aluminium in MONSTER modules.

\begin{figure}[htbp]
\begin{center}
\includegraphics[width=7.5cm, trim=0cm 0cm 0cm 0cm, clip=true]{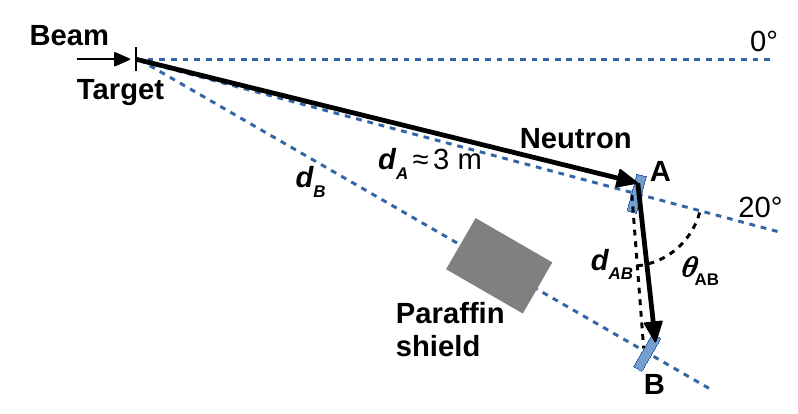}
\caption{Relative position of detectors $A$ and $B$, and of the paraffin shield in front of $B$. The distance $d_{AB}$ between the centres of the front faces of $A$ and $B$ and the corresponding scattering angle $\theta_{AB}$ from $A$ to $B$ are indicated. Only the liquid scintillator volumes of detectors $A$ and $B$ are shown.}
\label{Fig_detectors}
\end{center}
\end{figure}

One detector (\textit{detector} $A$) was placed at 20$^\circ$ from the charged particle beam axis, at around 3 m from the neutron production target, whilst the position of the second detector (\textit{detector} $B$) relative to $A$ was varied in order to measure the cross-talk probability at different neutron scattering angles.
Each detector was oriented with its entrance window facing the target and its axis crossing the target.
The positions of the detectors are characterised by the distance $d_{AB}$ between the centres of their front faces and the corresponding scattering angle $\theta_{AB}$ from the centre of the front face of $A$ to that of $B$ (Fig. \ref{Fig_detectors}).
Measurements were made at three different $\theta_{AB}$ angles: $\sim$50, $\sim$70 and $\sim$90$^\circ$.
The precise distances, $\theta_{AB}$ angles and angular ranges covered by $B$ are given in Table \ref{Tab_CT_conditions} together with the incident neutron energies and the types of detector modules.

Detector $B$ was shielded from directly incident neutrons by a paraffin block with a thickness of 60 cm and a cross section of 40 $\times$ 40 cm$^2$, to reduce the rate of $AB$ neutron-neutron coincidences resulting from two reactions in the same beam pulse. This shield was placed at 2 to 2.5 m from the target, depending on the relative position of $A$ and $B$.
MCNP simulations were performed to estimate its shielding efficiency (MCNP version 4C2 \cite{MCNP}).
At 6 and 16 MeV neutron energy, the neutron fraction that leaks through the shield with a residual energy above 600 keV ($\sim$ neutron energy threshold) is $\sim$6$\times10^{-5}$ and $\sim$5$\times10^{-3}$ respectively.
With this shield, given the beam intensities that were used, the reaction cross-sections, Poisson's statistics, the detector solid angles and the neutron detection efficiencies, the probability of coincident detection in $A$ and $B$ of two neutrons from the same beam pulse could be maintained at a negligible level (estimated to $<\!10^{-6}$ of the detection rate in $A$ or $<\!4\times 10^{-3}$ of the cross-talk probability).
The main neutron scattering sources in the vicinity of the detectors were the aluminium support structures, the steel and concrete walls, the concrete floor and the water-filled collimator and shield of a neutron intensity monitor.

The use of monoenergetic neutrons for cross-talk measurements offers several advantages. First, it allows for an unambiguous identification of the neutrons of interest incident on detector $A$.
In particular, neutrons from reactions on target impurities can be identified and rejected.
This was the case in the measurements with $\sim$4.8 MeV neutrons from the $D(d,n)^3$He reaction, where 1.5 MeV neutrons from the $(d,n)$ reaction on $^{12}$C impurities could be identified, as shown in the following.
In addition, with monoenergetic neutrons, the cross-talk events can be clearly identified since neutrons scattered from detector $A$ to $B$ are expected to have a well-defined $t_B - t_A$ TOF distribution that follows from the kinematics of the interaction in detector $A$. Consequently, uncorrelated events causing background are also easily identified, and fewer of these events are accepted in the selection of cross-talk events. Such events include, for example, ambient $\gamma$-rays, and $\gamma$-rays and scattered neutrons from neutron interactions on materials in the experimental hall.
To profit fully from these advantages requires the neutron TOF to be measured.
The charged particle beams were thus bunched at 2.5 MHz allowing for the generation of a ``start'' signal, which was produced by a beam pickup detector placed upstream of the neutron production target.
In most of the measurements, to improve the TOF resolution, the beam bunch duration was reduced from $\sim$10 ns to some 3-4 ns using Mobley magnetic compression \cite{Mobley}.
It should be noted that the bunch duration affected only the TOF resolution for neutrons incident on detector $A$, and not the TOF difference $t_B-t_A$ between $A$ and $B$ used to investigate the cross-talk events, as discussed below.
The absolute TOF calibration was obtained using the prompt $\gamma$-ray peak from the beam-target interactions as a reference.

\begin{figure}[htbp]
\begin{center}
\includegraphics[width=8cm]{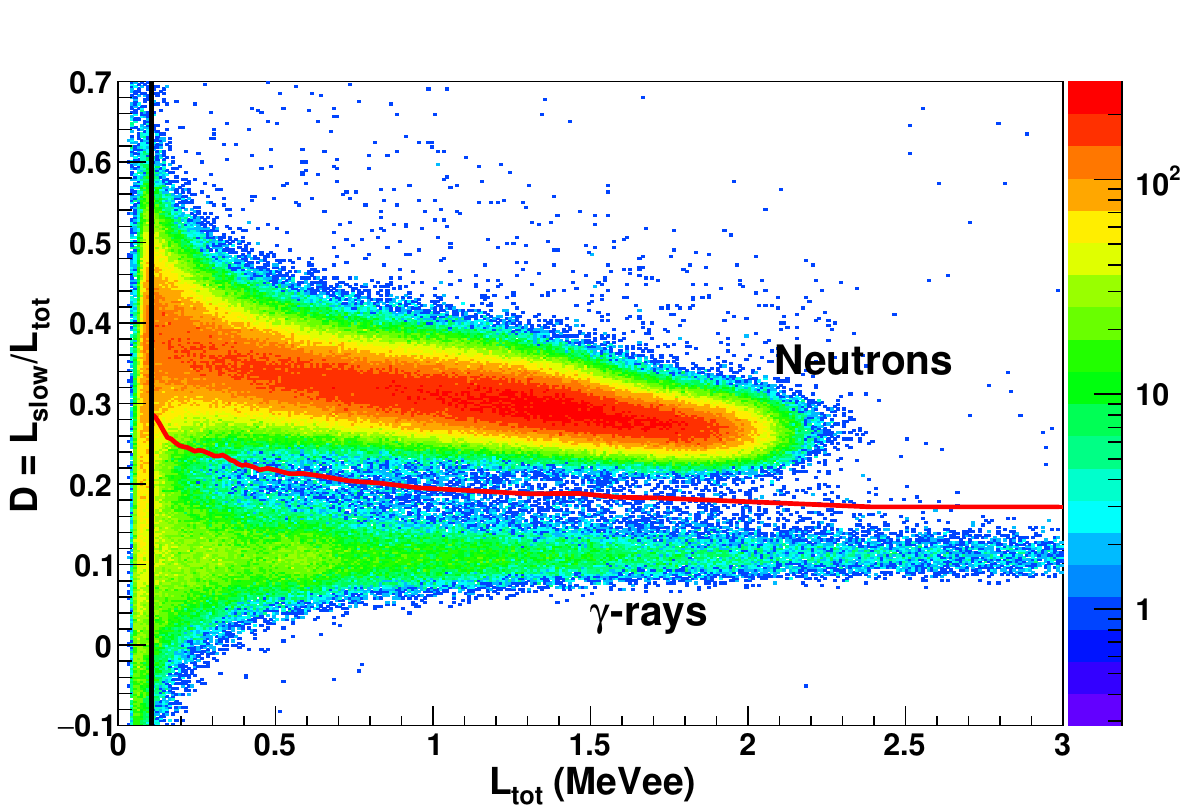}
\caption{Pulse-shape discrimination plot showing the ratio of the slow and total light outputs $L_{slow}/L_{tot}$ as a function of $L_{tot}$ in MeVee, in detector $A$ for $E_{n}=4.84$ MeV.
The red curve shows the limit used to select neutrons, whilst the vertical line shows the offline light output threshold $L_{thr}$ set at 100 keVee.
}
\label{Fig_PSD}
\end{center}
\end{figure}

Detectors $A$ and $B$ and the beam pickup detector were connected to the LPC-Caen FASTER digital electronics and data acquisition system \cite{FASTER}. Each channel uses a 12-bit, 500-MS/s digitizer with a 2-V input range, following an anti-aliasing low-pass filter limiting the input bandwidth to 100 MHz. The digitized signals are processed by an on-board FPGA that performs baseline subtraction, precise time stamping via a constant fraction discriminator for time measurement and signal integration over different time windows. Here, two windows were set on the signals from liquid scintillator detectors to integrate the total light output $L_{tot}$ and the ``slow'' light output $L_{slow}$.
The total integration window started 6 to 12 ns before the signal and had a duration of 250 ns, whilst the slow integration window was delayed by 40 ns and had a duration of 210 ns.
The calibration of $L_{tot}$ in terms of equivalent electron energy was performed using the full-energy peak of 60-keV $\gamma$-rays from a $^{241}$Am source and the Compton edge energies of $\gamma$-rays from $^{22}$Na, $^{137}$Cs, $^{88}$Y, $^{241}$Am-$^9$Be and $^{239}$Pu-$^{13}$C sources.
A second-order polynomial was adjusted to the points (electron energy vs $L_{tot}$) to determine the calibration function.
The two light integrals $L_{tot}$ and $L_{slow}$ were used to perform neutron-$\gamma$ pulse-shape discrimination (PSD) via the charge comparison method based on the relation between $L_{slow}$ and $L_{tot}$.
More precisely, the $L_{slow}/L_{tot}$ ratio was used as the discrimination variable $D$ and plotted as a function of $L_{tot}$ to select neutrons.
A detailed study has shown the excellent neutron-$\gamma$ PSD performance of this digital system with organic scintillators \cite{Senoville}.
An example of pulse shape discrimination plot is shown in Fig. \ref{Fig_PSD} for detector $A$ and $E_n=4.84$ MeV, together with the cut used to select neutrons and the light output threshold set on $L_{tot}$.
The pulse-shape discrimination quality at a given light output $L_{tot}$ can be measured by the figure of merit \cite{BrooksFoM} $M = (D_n-D_\gamma)/(W_n+W_\gamma)$, where $D_{n}$ and $W_n$ are, respectively, the most probable value and full width at half maximum of the neutron distribution at $L_{tot}$, whilst $D_\gamma$ and $W_\gamma$ are the corresponding quantities for $\gamma$-rays.
For our detectors, $M = 1.65$ at $L_{tot}=1$ MeVee and $M=1.0$ at $L_{tot} = 0.23$ MeVee.

\begin{figure}[htbp]
\begin{center}
\includegraphics[width=9.5cm]{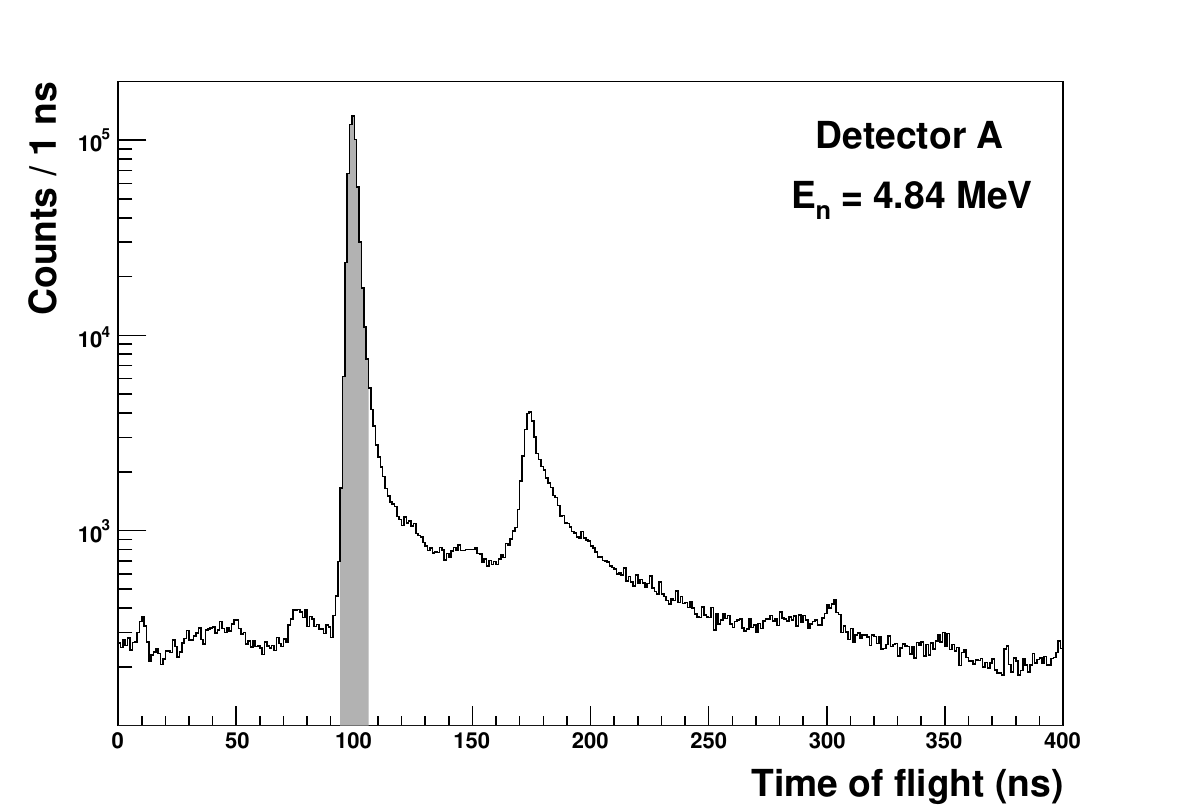}
\caption{Time-of-flight spectrum measured for neutron events selected by PSD in detector $A$, in the case of 4.84 MeV neutron production with the D$(d,n)^3$He reaction. The shaded area shows the time-of-flight gate applied to select the 4.84 MeV incident neutrons.}
\label{Fig_TOF_A}
\end{center}
\end{figure}

Fig. \ref{Fig_TOF_A} shows an example of calibrated time-of-flight spectrum for neutron events selected by PSD in detector $A$, with 4.84 MeV neutrons produced by the D$(d,n)^3$He reaction. The peak at 99 ns corresponds to the 4.84 MeV neutrons, whilst peaks at TOF $>$ 150 ns are due to reactions on target impurities. In particular, the peak at $\sim$175 ns corresponds to 1.55 MeV neutrons produced by the $(d,n)$ reaction on $^{12}$C impurities in the target. The weak peak at $\sim$10 ns corresponds to prompt $\gamma$-rays from the target.

\section{Results and discussion}

The following conditions were employed to select the cross-talk events.
\begin{itemize}
\item TOF of detector $A$ corresponding to a direct neutron from the target with the energy of interest (Table \ref{TabBeamsTargets}). An example of TOF gate is shown in Fig. \ref{Fig_TOF_A}.
\item Events detected in coincidence between detectors $A$ and $B$.
\item Events in detectors $A$ and $B$ identified as neutrons by PSD, with the total light output $L_{tot}$ above a software threshold $L_{thr}$ set to ensure good PSD. A threshold of 100 keVee was employed for both detectors $A$ and $B$ (630 keV neutron energy using the light-response function for protons in NE213 of Cecil et al. \cite{Cecil}), except for $E_n = 1.42$ MeV where the threshold was lowered to 75 keVee (530 keV neutron energy) to detect lower energy cross-talk neutrons.
\end{itemize}

% EXP RESULTS

\begin{figure}[htbp]
\begin{center}
\includegraphics[width=8cm]{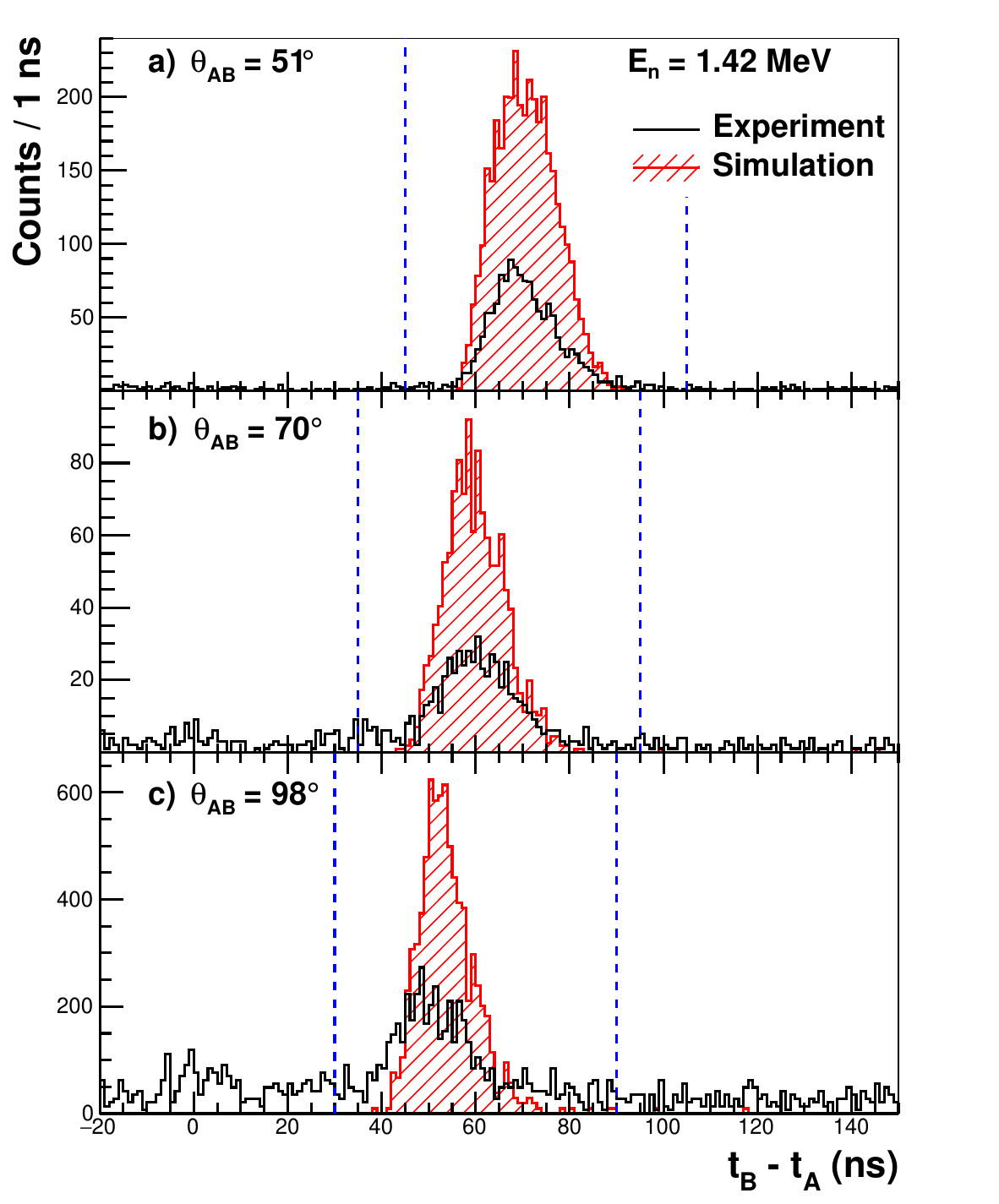}
\caption{Distributions of the $t_B-t_A$ time-of-flight for $E_n = 1.42$ MeV, and $\theta_{AB} = 51 ^\circ$ (a), $70 ^\circ$ (b) and $98 ^\circ$ (c). The black histograms show the experimental data after cross-talk selection (see text for details), whilst the red hatched histograms show the results of GEANT4 simulations including the scintillator volumes, the cell walls and glass windows, the light guides and the external casings of detectors $A$ and $B$. The blue dashed vertical lines indicate the integration range of the experimental spectrum for the determination of the cross-talk probability.}
\label{TOF_AB_1MeV42}
\end{center}
\end{figure}

\begin{figure}[htbp]
\begin{center}
\includegraphics[width=8cm]{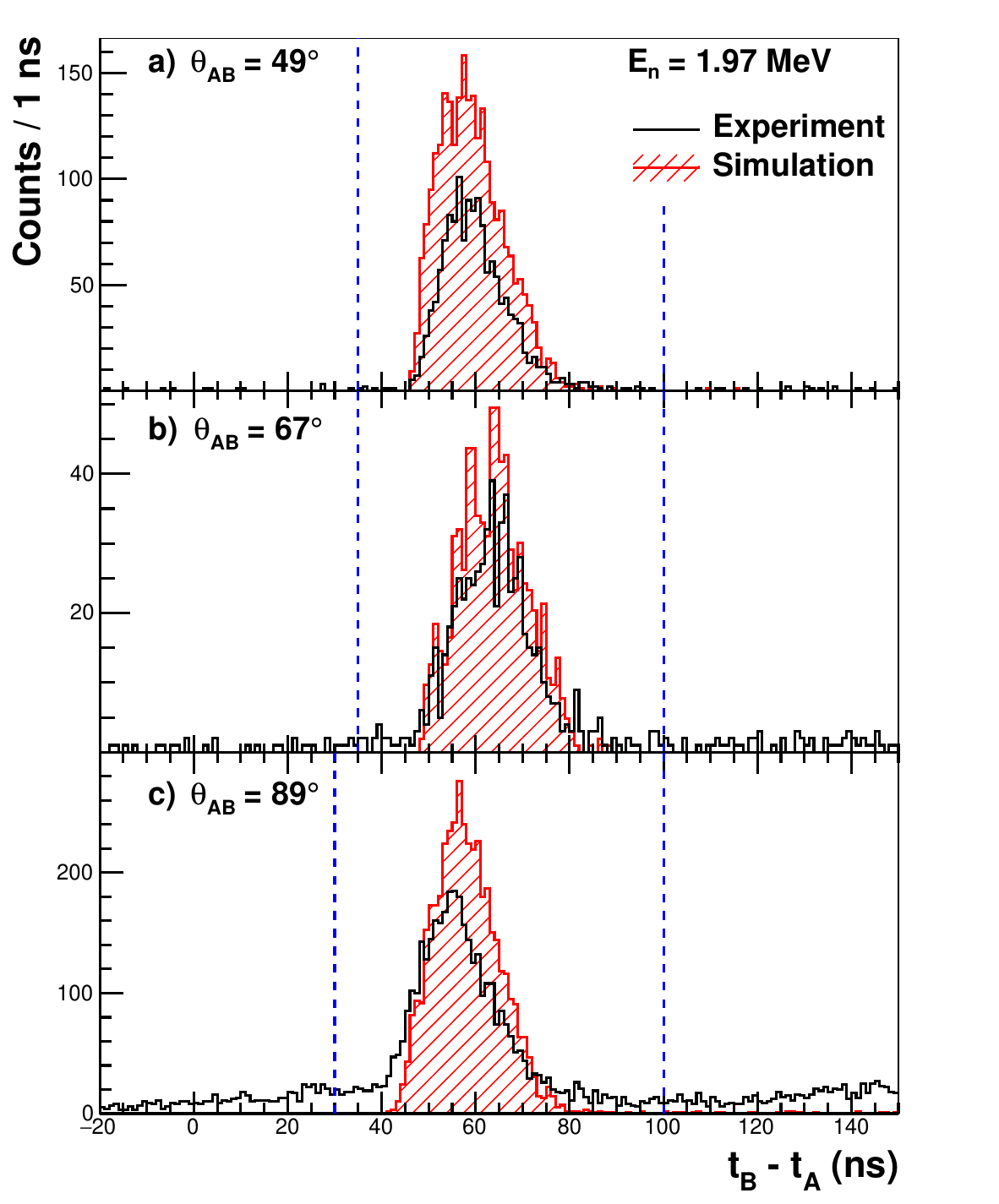}
\caption{Same as Fig. \ref{TOF_AB_1MeV42} for $E_n = 1.97$ MeV, and $\theta_{AB} = 49 ^\circ$ (a), $67 ^\circ$ (b) and $89 ^\circ$ (c).}
\label{TOF_AB_1MeV97}
\end{center}
\end{figure}

\begin{figure}[htbp]
\begin{center}
\includegraphics[width=8cm]{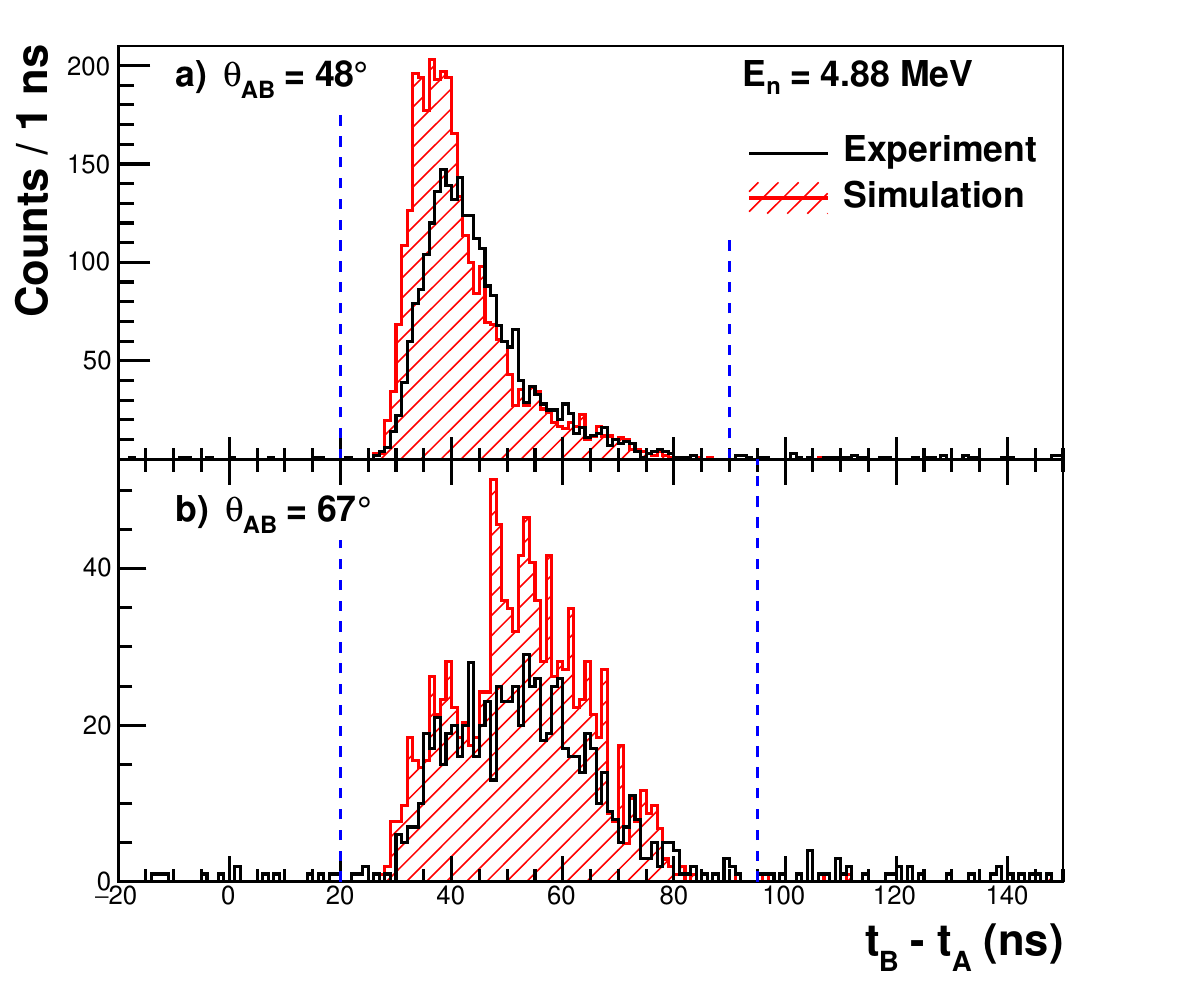}
\caption{Same as Fig. \ref{TOF_AB_1MeV42} for $E_n = 4.88$ MeV, and $\theta_{AB} = 48 ^\circ$ (a) and $67 ^\circ$ (b).}
\label{TOF_AB_4MeV88}
\end{center}
\end{figure}

\begin{figure}[htbp]
\begin{center}
\includegraphics[width=8cm]{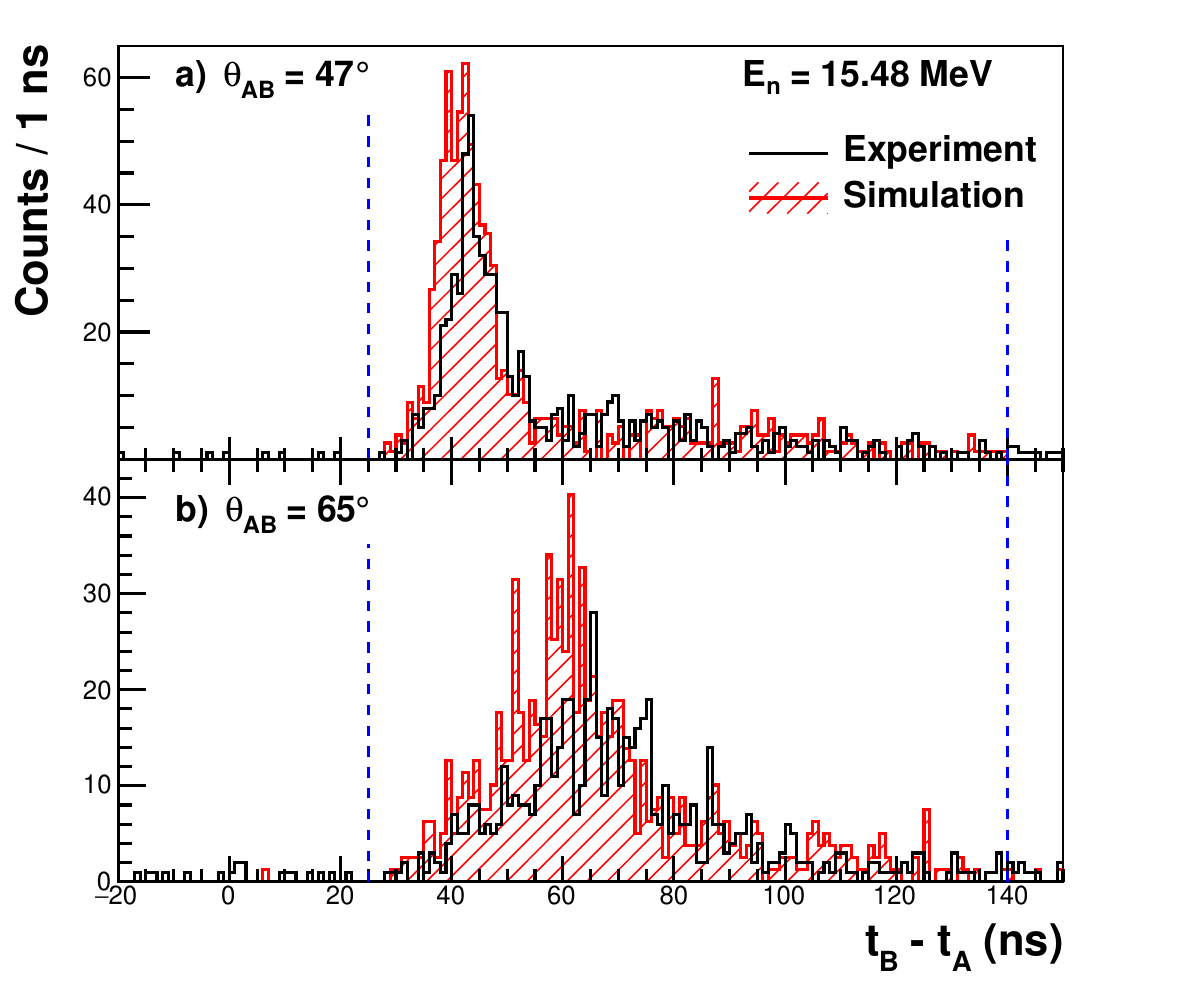}
\caption{Same as Fig. \ref{TOF_AB_1MeV42} for $E_n = 15.48$ MeV, and $\theta_{AB} = 47 ^\circ$ (a) and $65 ^\circ$ (b).}
\label{TOF_AB_15MeV48}
\end{center}
\end{figure}

Figs. \ref{TOF_AB_1MeV42}-\ref{TOF_AB_15MeV48} present the $t_B - t_A$ TOF spectra for the events selected with the above conditions, for incident neutron energies of 1.42, 1.97, 4.88 and 15.48 MeV and the different $\theta_{AB}$ angles.
The spectra for the data taken at 2.31 MeV are similar to those at 1.97 MeV, whilst those measured at 4.84 MeV are almost identical to the 4.88 MeV data.
A full listing of the data sets and experimental conditions is provided in Table \ref{Tab_CT_conditions}.
It should be noted that the ratio of the numbers of counts in two data sets at different energies and/or angles does not reflect the ratio of the cross-talk rates, since the beam intensity and the acquisition time were adapted to each measurement.
The experimental distributions of Figs. \ref{TOF_AB_1MeV42}-\ref{TOF_AB_15MeV48} have a simple structure, exhibiting a single, rather well-defined peak in the $t_B-t_A>0$ region, over a range of a few tens of ns, arising from cross-talk. A significant background is visible only at the lowest neutron incident energies and at large angles ($\theta_{AB} \sim 90^\circ$) where the cross-talk rates are the smallest.

\begin{figure}[htbp]
\begin{center}
\includegraphics[width=8cm]{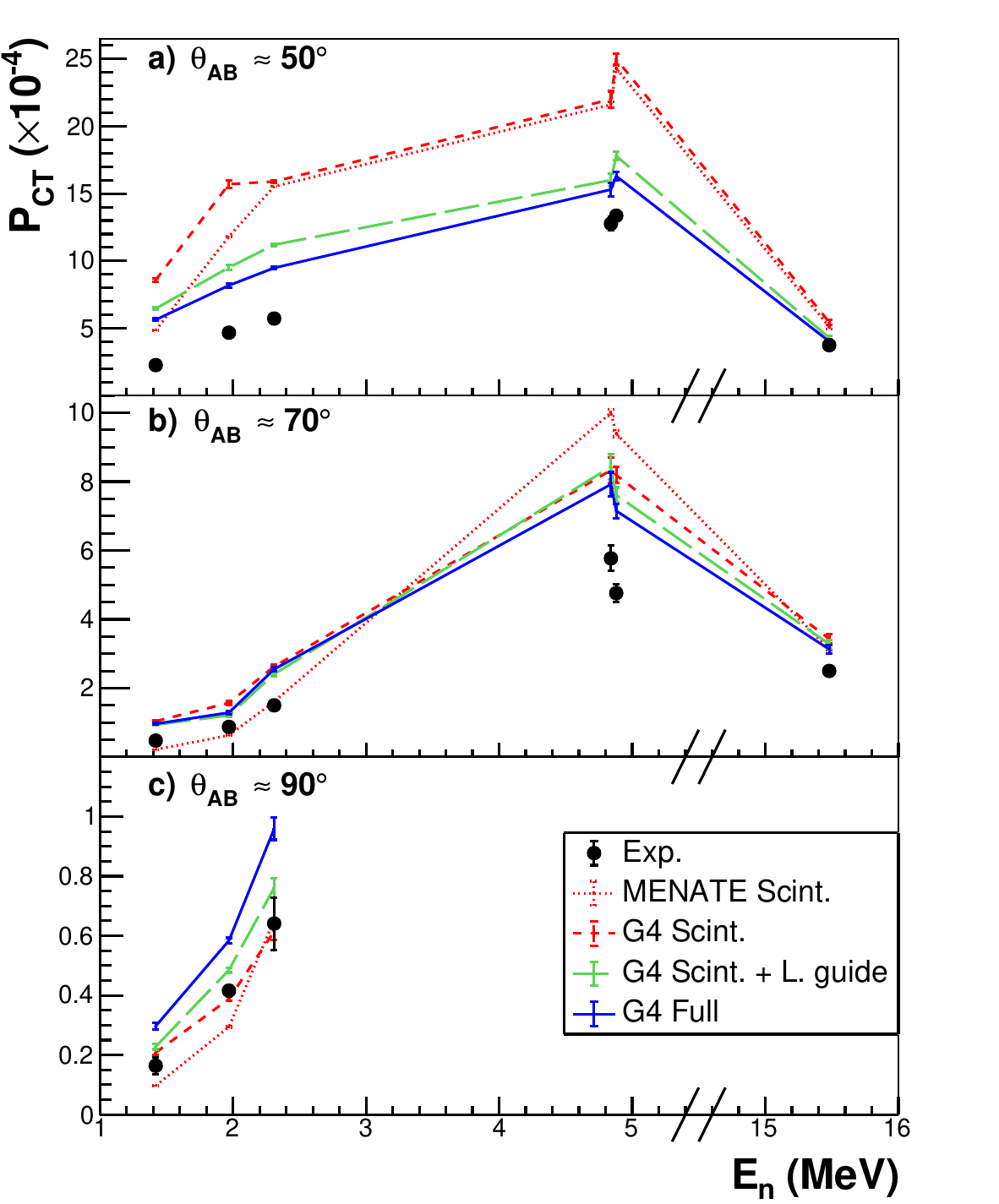}
\caption{Experimental cross-talk probabilities as a function of neutron incident energy, compared to results from simulations performed with MENATE and GEANT4 (``G4''), with only the scintillator volume (``Scint.''), the scintillator volume and light guide (``Scint + L. guide''), and with the scintillator volume, scintillator cell, glass window, light guide and casing materials (``Full''). The lines connecting the simulation results are drawn to guide the eye. Error bars are statistical and, for the experimental results, include the contribution from background subtraction.}
\label{PCT_vs_E}
\end{center}
\end{figure}

To measure the number of cross-talk events for a given $E_n$-$\theta_{AB}$ configuration, the experimental peak was integrated between the bounds indicated by vertical dashed lines on Figs. \ref{TOF_AB_1MeV42}-\ref{TOF_AB_15MeV48}. These bounds were set where the data reach the background level.
The background was averaged over several channels (typically 10) around each bound and subtracted assuming a linear dependence on $t_B~-~t_A$.
The experimental cross-talk probability $P_{CT,exp}$ was defined as the ratio of the number of cross-talk events $N_{AB}$ and the number of neutrons $N_A$ detected in $A$ with the incident energy of interest.
Table \ref{Tab_PCT} lists the measured cross-talk probabilities for each $E_n$-$\theta_{AB}$ combination.
They are also shown in Fig. \ref{PCT_vs_E} as a function of $E_n$, for the three $\theta_{AB}$ angles.
The uncertainties on the cross-talk probabilities are statistical and include a contribution from background determination and subtraction.
Globally, the cross-talk probability decreases with the $\theta_{AB}$ angle for a given neutron energy, and increases with energy up to 4.88 MeV.
At 15.48 MeV the cross-talk rates are reduced compared to 4.88 MeV.

\begin{figure}[htbp]
\begin{center}
\includegraphics[width=8cm]{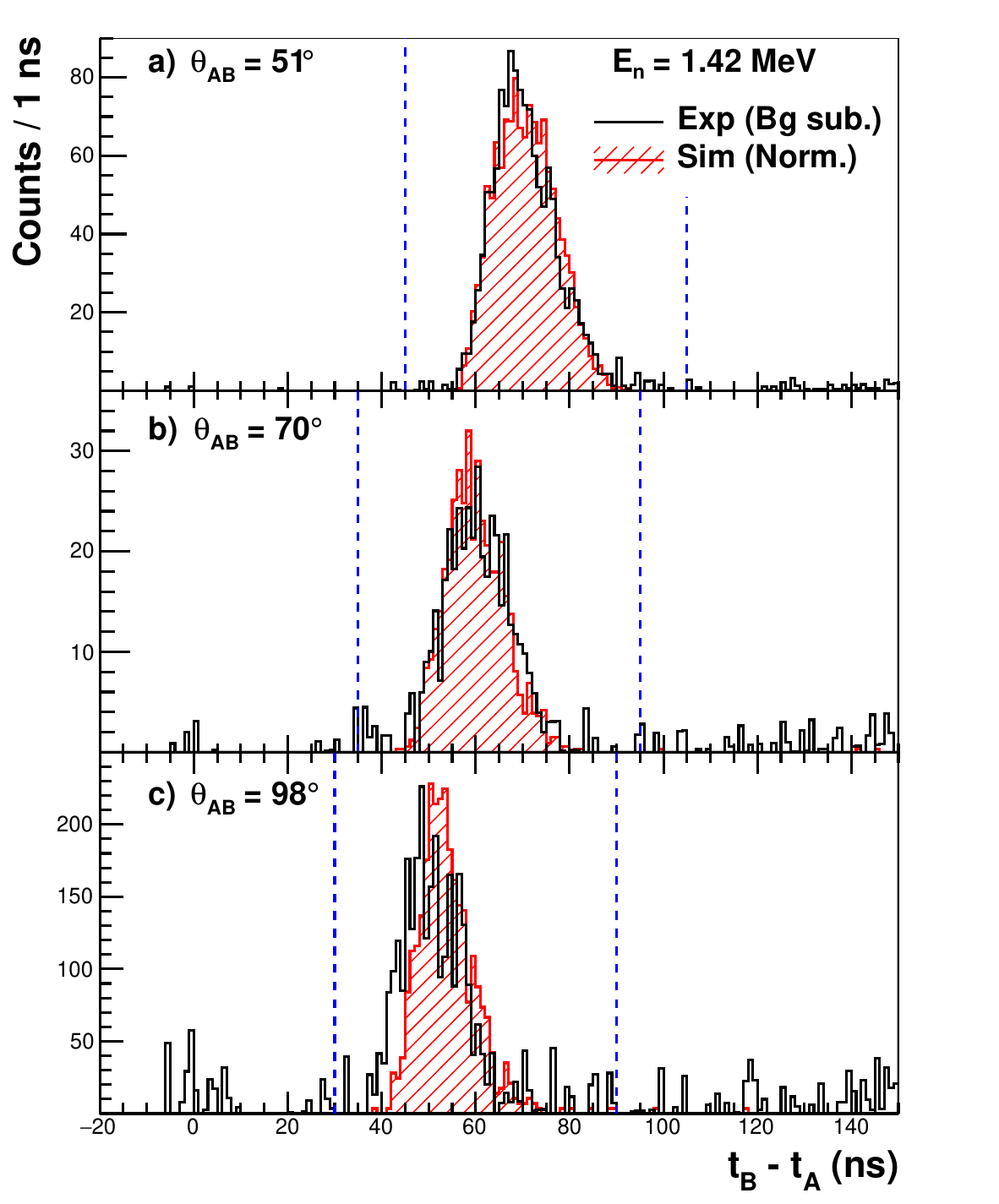}
\caption{Distributions of the $t_B-t_A$ time-of-flight for $E_n = 1.42$ MeV, and $\theta_{AB} = 51 ^\circ$ (a), $70 ^\circ$ (b) and $98 ^\circ$ (c). The black histograms show the experimental data after cross-talk selection (see text for details) and background subtraction.
The blue dashed vertical lines indicate the integration range of the experimental spectrum for the determination of the cross-talk probability.
The red hatched histograms show the results of GEANT4 simulations including the scintillator volumes, the cell walls and glass windows, the light guides and the external casings of detectors $A$ and $B$, normalized to the same integral between the indicated bounds as in the background-corrected experimental spectrum.}
\label{TOF_AB_1MeV42_Sub_Norm}
\end{center}
\end{figure}

\begin{figure}[htbp]
\begin{center}
\includegraphics[width=8cm]{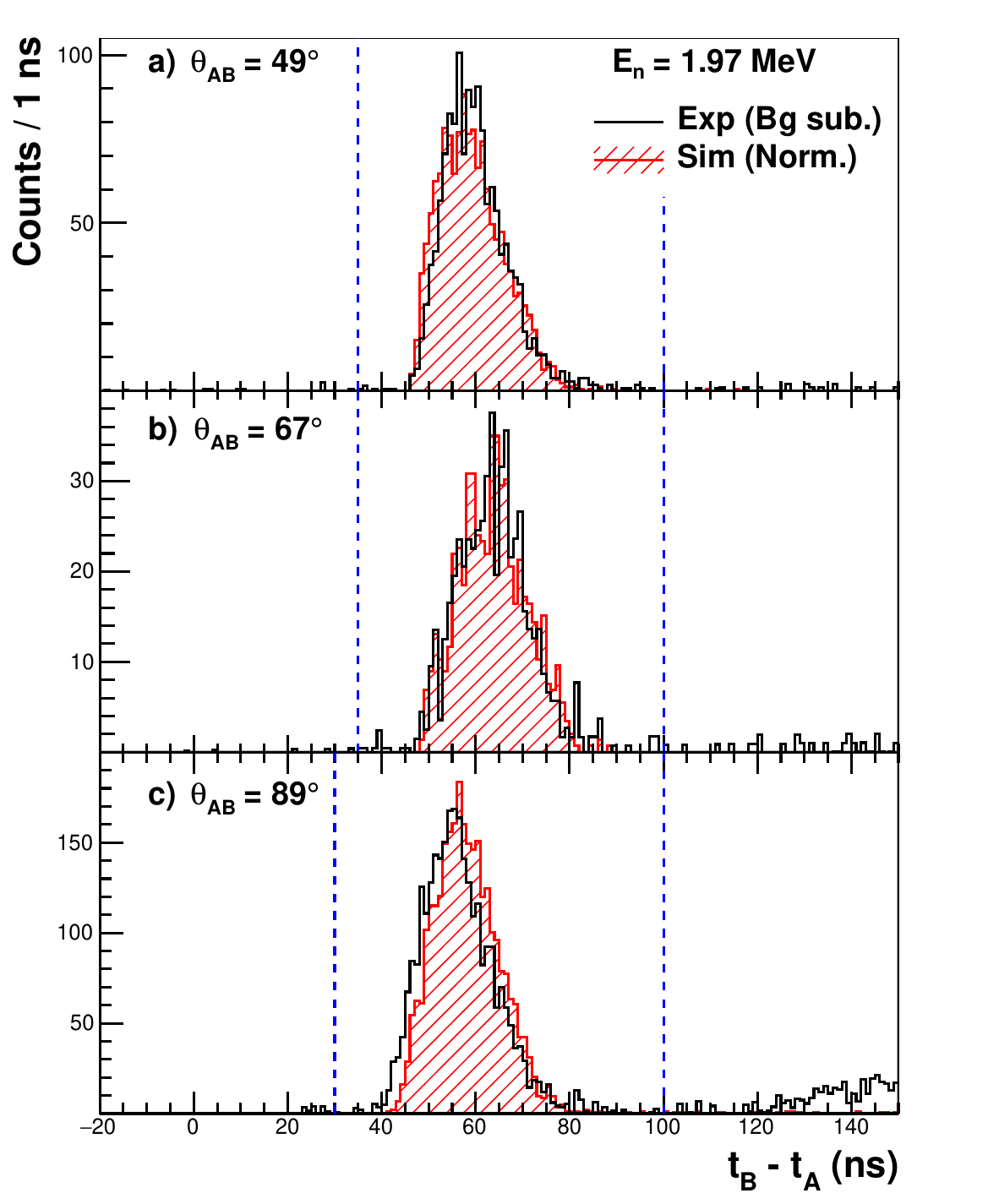}

\caption{Same as Fig. \ref{TOF_AB_1MeV42_Sub_Norm} for $E_n = 1.97$ MeV, and $\theta_{AB} = 49 ^\circ$ (a), $67 ^\circ$ (b) and $89 ^\circ$ (c).}
\label{TOF_AB_1MeV97_Sub_Norm}
\end{center}
\end{figure}

The experimental $t_B-t_A$ TOF spectra are compared in Figs. \ref{TOF_AB_1MeV42}-\ref{TOF_AB_15MeV48} with results of simulations performed with GEANT4 \cite{GEANT4}, including the scintillator volume and the surrounding inactive materials, i.e. the walls of the scintillator cell, the glass window on the PMT side of the cell, the light guide and the external casing of the detectors.
The compositions and densities of the various materials included in the simulations are given in Table \ref{Tab_Materials}.
The ``Shielding'' Physics List that incorporates, in addition to electromagnetic and hadronic processes, the ``NeutronHP'' model for the interactions of neutrons was used. The latter is based on evaluated neutron-induced reaction cross section and angular distribution data and extends up to 20 MeV neutron energy. The GEANT4 version 4.10.06 was employed, in which the NeutronHP model offers as an option an improved description of the $^{12}\text{C}(n,\alpha)^9\text{Be}$ and $^{12}\text{C}(n,n^\prime3\alpha)$ reaction channels \cite{Garcia}. These two channels open at 6.2 and 8.8 MeV neutron energy, respectively, and can potentially affect the simulations for the 15.5 MeV measurements.
Light output quenching was included in the post-processing of the simulation results, via energy-to-light functions \cite{Cecil,Batchelor}. These functions were used to convert the energy deposited by each secondary charged particle into light output. The total light output was then compared to a threshold set to the same value as in the analysis of the experimental data.
The effects of PSD were also included by requiring that more than 50 \% of the total light output be produced by heavy charged particles for the event to be considered as a neutron event, the remaining light being produced by electrons from $\gamma$-ray interactions.
For a given $E_n$-$\theta_{AB}$ configuration, the simulated distribution was obtained with the total number of incident neutrons detected in $A$ equal to that measured experimentally for the energy of interest.
Therefore the ratio of the experimental and simulated numbers of cross-talk events reflects the ratio of cross-talk probabilities.
For each configuration, the position and shape of the simulated distribution are in very good agreement with those of the main experimental peak.
Even in cases where the experimental peak is clearly asymmetric (at $E_n = 4.88$ MeV and $\theta_{AB} = 48^\circ$, and at 15.48 MeV) the agreement is very good.
At low energy, the simulations overestimate the cross-talk rates, and the background at $\theta_{AB}\sim90^\circ$ becomes significant.
For a clearer comparison of the experimental and simulated distributions, Figs. \ref{TOF_AB_1MeV42_Sub_Norm} and \ref{TOF_AB_1MeV97_Sub_Norm} show the experimental spectra after background subtraction and the simulations normalized to the same peak integral as in the experimental data, between the bounds indicated by the dashed vertical blue lines.
At low energy and $\theta_{AB}\sim90^\circ$, the simulated cross-talk peak is shifted by a few ns towards higher $t_B - t_A$ compared to the experimental peak, whilst its shape agrees well with the data. The shift is smaller at 1.97 MeV than at 1.42 MeV. A similar shift also exists at 2.31 MeV and $\theta_{AB}=98^\circ$, but is also smaller than at 1.97 MeV.
As discussed below, the cross-talk at low energy and large angles is mainly due to multiple interactions in detector $A$, in particular involving $n$+C scattering. As such, it is most likely that these multiple scattering processes are not fully accounted for in the simulations.

Table \ref{Tab_PCT} and Fig. \ref{PCT_vs_E} compare the experimental and simulated cross-talk probabilities.
In addition to the GEANT4 simulations including the scintillator volume and surrounding materials discussed above, simulations were performed with fewer inactive detector elements included.
The simplest simulations were performed with only the scintillator, using GEANT4 and the MENATE code \cite{Desesquelles}.
The latter includes neutron-proton ($np$) and neutron-$^{12}$C elastic scattering and a set of neutron-induced reactions on $^{12}$C. As such, it allows for the description of interactions in the scintillator only.
In a third series of GEANT4 simulations, the scintillator and light guide were included.
The uncertainties on the simulated cross-talk probabilities are statistical.
The simulations including only the scintillator volumes with MENATE and GEANT4 overestimate the cross-talk rate at $\sim$50° and $E_n \lesssim 5$ MeV by a factor of $\sim$2 to $\sim$4. This discrepancy is reduced as the angle increases.
The two simulation codes show a significant disagreement with each other at the lowest energies.
Inclusion of the light guide in the GEANT4 simulations strongly reduces the cross-talk probability at all energies at $\sim$50° and as such greatly improves the agreement with the data.
Indeed, at this angle, a large fraction of the neutrons scattered in detector $A$ must traverse the light guide of $A$ without interacting to be able to reach $B$ and produce cross-talk.
The effect is also visible at $\sim$70°, but somewhat smaller.
At $\sim$90° the presence of the light guide increases the cross-talk rate, and slightly degrades the agreement with the data.
The importance of the guide in terms of neutron interactions stems from its dimensions, density and composition (PMMA, which contains H, C and O) similar to those of the scintillator.
As such, it can lead to the rescattering of the neutrons, thus influencing cross-talk.
Adding the other inactive materials further improves the agreement with the data throughout the energy range at $\sim$50°.
At $\sim$70°, the effect is similar for $E_n \gtrsim 5$ MeV, but is very small for lower energies.
On the contrary, at $\sim$90° and $E_n<2.5$ MeV, the additional materials increase $P_{CT,sim}$ and the agreement worsens.
At these low energies and large effective scattering angles, single neutron scattering can not produce cross-talk, as discussed below.
The cross-talk probability is then very small and likely more sensitive to details of the material description, e.g. a more complex composition or impurities, beyond the ideal materials assumed in the simulations.
A consistent overestimation of the cross-talk by the ``full'' simulations at all energies and angles is observed, typically smaller than 40 \% but reaching a factor $\sim$2 at $E_n = 1.42$ MeV.
Nevertheless, the overall evolution of the experimental cross-talk probability with energy and angle, with a decrease by one order of magnitude from $\sim$50$^\circ$ to $\sim$90$^\circ$, is well reproduced by the simulations.

The larger discrepancy between the experimental cross-talk probabilities and the simulations at the lowest energies ($E_n<2.5$ MeV) calls for some comments. At these energies, the neutrons detected in detector $A$ and scattered towards detector $B$ have energies of order $\sim$1-2 MeV, where the cross sections are usually higher than at $E_n \gtrsim 5$ MeV and can present resonances. In addition, the amount of scintillation light produced is low. The simulations then become more sensitive to inputs such as the material compositions, densities or thicknesses, the light output functions, or the light output threshold, which all are subject to uncertainties. Cross-talk is particularly critical as it involves two detectors and, as discussed below, multiple scattering for some detector configurations, whereby the effects of such uncertainties are enhanced. This might be one of the factors explaining why adding the inactive materials, whose characteristics are not known very precisely, does not improve the agreement with the data at all angles. The sensitivity to the light output calibration or to the light output functions can be evaluated by modifying the light output threshold in the simulations. For example, at $E_n=1.42$ MeV, changing the threshold by 7 \% leads to a change in the cross-talk probability of 11 \% at $\theta_{AB}=51^\circ$, 20 \% at 70$^\circ$ and 26 \% at 98$^\circ$. An indication that substantial uncertainties affect the simulations is the significant difference observed at low energies between the cross-talk probabilities predicted by the MENATE and GEANT4 simulations including only the scintillator volume.

In the present energy range, the main interaction process involved in neutron detection is $np$ elastic scattering, given its efficient energy transfer to the proton and the high light output of the scintillator for protons compared to heavier recoils \cite{Cecil,Verbinski}.
For some of the $E_n$-$\theta_{AB}$ configurations used here, a single $np$ scattering in $A$ can not produce a cross-talk event, either because the necessary neutron scattering angle is larger than the maximum possible angle from $np$ scattering ($\sim$88$^\circ$ given the neutron-to-proton mass ratio), or because the energy of the scattered neutron is below the neutron energy detection threshold.
This is the case for the configurations with $\theta_{AB} \sim 90^\circ$, and for $\theta_{AB} = 70^\circ$ at $E_n = 1.42$ MeV.
However, our measurements and simulations show that cross-talk can occur in these cases.
It is therefore interesting to investigate the origin of cross-talk events, which can be done using the simulations. In the following, we consider results from the ``full'' GEANT4 simulations that include, in addition to the scintillator volume, the scintillator cell walls, glass window, light guide and casing of the detectors.

\begin{figure}[htbp]
\begin{center}
\includegraphics[width=8cm]{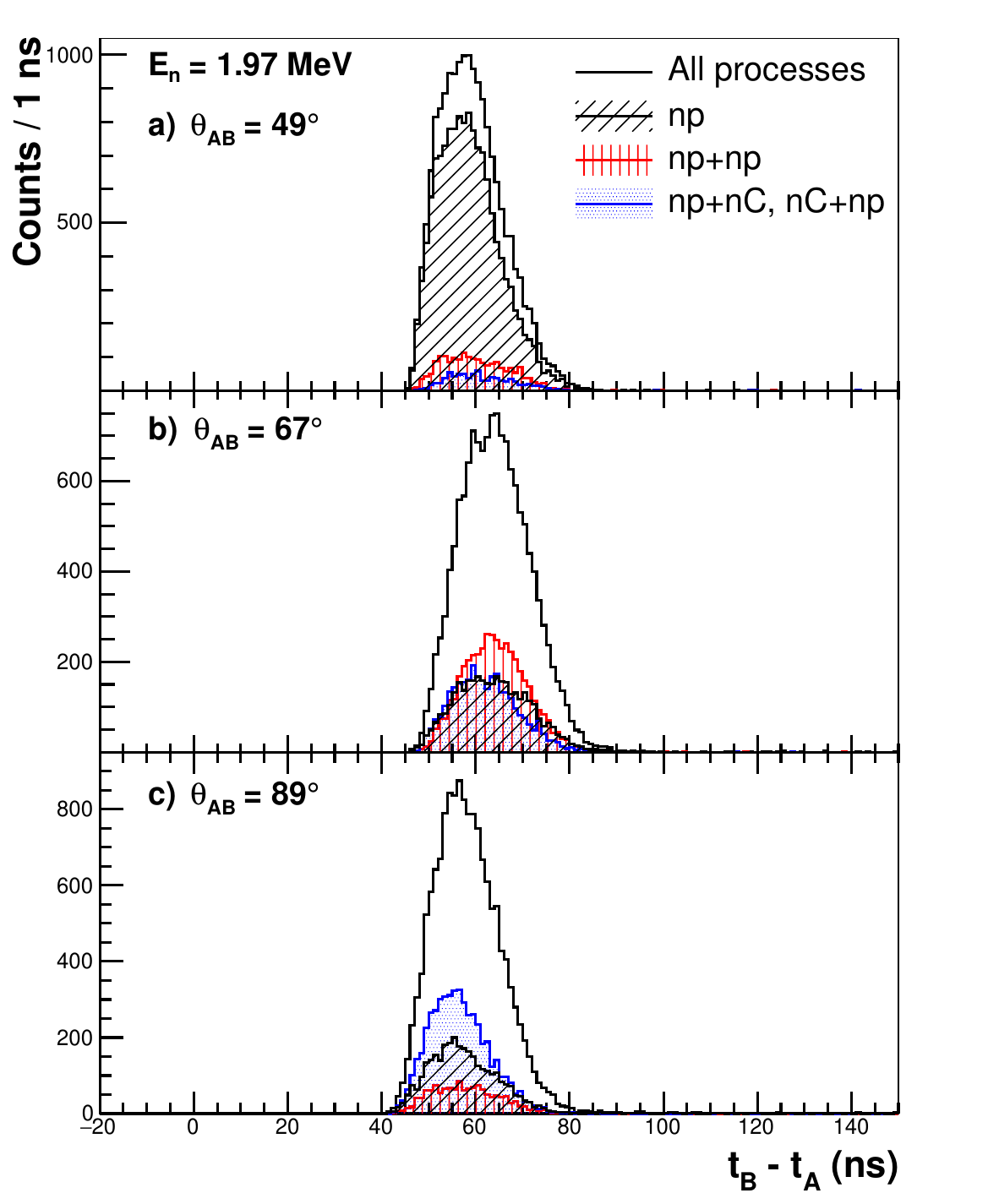}
\caption{Simulated $t_B-t_A$ distribution for the cross-talk events at 1.97 MeV, for $\theta_{AB} = 49^\circ$ (a), 67$^\circ$ (b) and 89$^\circ$ (c). The various histograms show the main neutron interaction processes in detector $A$ responsible for cross-talk.}
\label{TOF_AB_1MeV97_SimProcesses}
\end{center}
\end{figure}

\begin{figure}[htbp]
\begin{center}
\includegraphics[width=8cm,clip=true,trim=0 0 0 5.7cm]{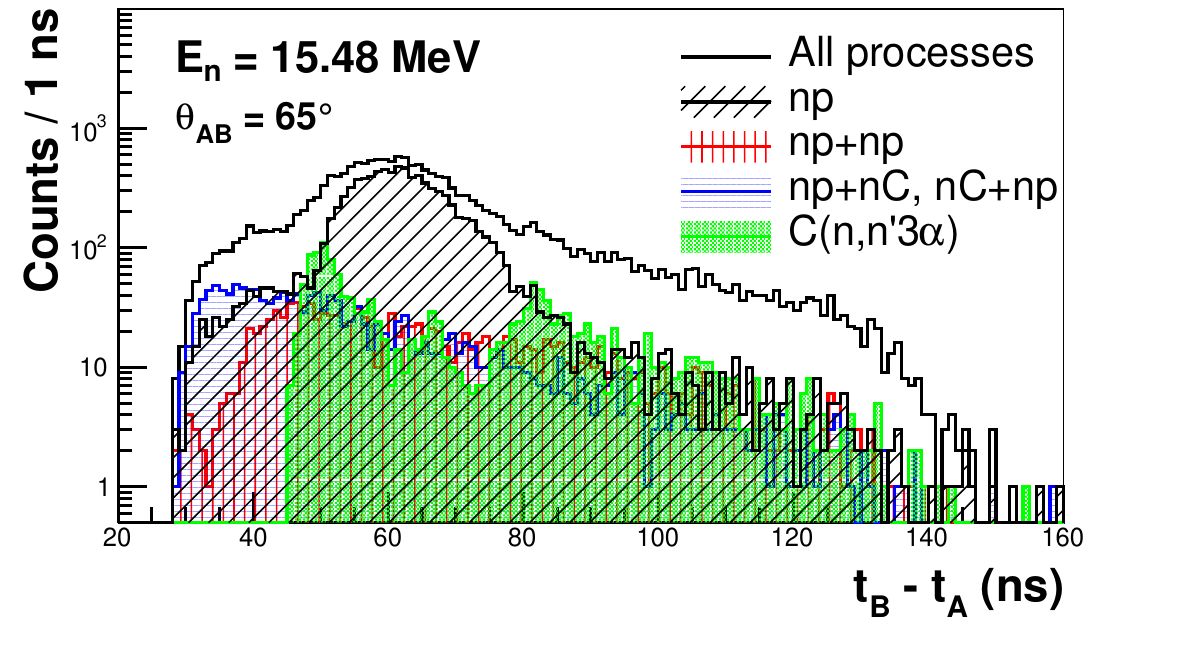}
\caption{Simulated $t_B-t_A$ distribution for the cross-talk events at 15.48 MeV and $\theta_{AB} = 65^\circ$. The various histograms show the main neutron interaction processes in detector $A$ responsible for cross-talk.}
\label{TOF_AB_15MeV48_SimProcesses}
\end{center}
\end{figure}

The contributions of the main processes in detector $A$ to the simulated cross-talk events are given in Table \ref{Tab_SimProcesses} for $E_n$=1.97, 4.84 and 15.48 MeV. The corresponding $t_B-t_A$ TOF spectra for 1.97 and 15.48 MeV are shown on Figs. \ref{TOF_AB_1MeV97_SimProcesses} and \ref{TOF_AB_15MeV48_SimProcesses}, respectively.
At  $E_n = 1.97$ MeV and $\theta_{AB} \sim 50^\circ$ (see Fig. \ref{TOF_AB_1MeV97_SimProcesses}a), cross-talk from a single $np$ scattering in $A$ is kinematically allowed. In this case, the simulations indicate that most of the cross-talk events originate from this process, whilst double elastic scattering interactions, $np+np$, $np+n$C and $n\text{C}+np$, are responsible for some 17 \% of the events.
In these cases, $n$C scattering plays a role by allowing for a large-angle scattering towards $B$ without a significant neutron energy loss, whilst $np$ scattering ensures enough scintillation light production in $A$ for detection.
For the $\theta_{AB} \sim 70^\circ$ configuration, a unique $np$ collision in $A$ can produce cross-talk for only a very narrow range of neutron scattering angles ($\sim[54^\circ,56^\circ]$).
Fig. \ref{TOF_AB_1MeV97_SimProcesses}b and Table \ref{Tab_SimProcesses} show that in this case single $np$ scattering is not the dominant process and that double scattering events contribute to a very large fraction of the cross-talk events (53 \%). Triple scattering processes also become non negligible.
For $\theta_{AB} \sim 90 ^\circ$, the most probable mechanism is $np+n$C or $n\text{C}+np$ double scattering, whilst triple scattering involving $n$C is also significant.
At this angle and energy, in principle neutrons from single $np$ scattering do not have enough energy to be detected in $B$.
The presence of cross-talk events from single $np$ scattering in $A$, as seen on Fig. \ref{TOF_AB_1MeV97_SimProcesses}c and Table \ref{Tab_SimProcesses}, is due to rescattering in the light guide. In simulations without the light guide, these events indeed disappear.
The simulations indicate that the cross-talk mechanisms at 1.42 and 2.31 MeV and a given $\theta_{AB}$ angle are similar to those at 1.97 MeV at the corresponding angle. For the other incident energies, 4.84, 4.88 and 15.48 MeV, at $\theta_{AB} \sim 50^\circ$ and 70$^\circ$ cross-talk from a single $np$ scattering in $A$ is possible and is the dominant process. However, double scattering remains a significant source of cross-talk.

The $E_n$ = 15.48 MeV case deserves further attention as other neutron reaction channels are open. In addition to single $np$ and double scattering in detector $A$, which are responsible for 50-60 \% and 17 \% of the cross-talk events, respectively, cross-talk can originate from a $^{12}\text{C}(n,n^\prime3\alpha)$ reaction in $A$ ($\sim$ 10 \% of the events). However, the other open reaction channels, $^{12}\text{C}(n,\alpha)^9\text{Be}$, $^{12}\text{C}(n,p)^{12}\text{B}$ and $^{12}\text{C}(n,d)^{11}\text{B}$, without a residual neutron, can not be sources of cross-talk when they occur in $A$.
As shown in Fig. \ref{TOF_AB_15MeV48_SimProcesses} for $\theta_{AB} = 65^\circ$, the $t_B-t_A$ spectrum at $E_n$ = 15.48 MeV is more complex.
It is characterized by a main peak with a tail towards higher TOF. The simulations show that the peak is mainly due to single $np$ scattering, whilst the $^{12}\text{C}(n,n^\prime3\alpha)$ reaction and double scattering mechanisms give a much broader $t_B-t_A$ distribution and contribute to a large fraction of the cross-talk events in the tail.

Since $np$ scattering is the dominant interaction process for neutron detection, it is customary practice for cross-talk identification filters, for relatively low energy neutrons ($E_n \lesssim 30$ MeV), to assume single $np$ scattering as the origin of cross-talk \cite{Wang, MarquesCT}.
However, for all neutron energies and geometrical configurations considered here, our simulations indicate the presence of a significant fraction of cross-talk events (at least $\sim$ 20 \% and up to $\sim$ 80 \%) due to other mechanisms, such as multiple scattering.
This will limit the efficiency of cross-talk filters that assume single $np$ scattering.
Filters based on more general kinematical criteria will have to be applied, such as conditions based on the relative times and distances between the detectors, and on the energies deposited in the detectors deduced from the light outputs without assuming proton recoils, employed previously to identify cross-talk induced by $E_n \sim 30$ MeV neutrons \cite{Pluta,MarquesCT}.

\section{Summary and conclusion}

A series of measurements to investigate the characteristics of neutron cross-talk between organic scintillator modules at low energies have been performed with monoenergetic neutrons. Cross-talk time-of-flight spectra and probabilities were determined for neutron energies from 1.4 to 15.5 MeV and effective scattering angles from $\sim$50$^\circ$ to 100$^\circ$. These were compared to a range of simulations including both the scintillator volumes and the inactive materials of the detector modules.

The simulations were found to reproduce very well the cross-talk time-of-flight spectra whilst the cross-talk probabilities were in approximate agreement with the measurements -- typically within 40 \% or better, with the most significant discrepancies occurring at the lowest neutron energies. The inclusion of the light guides in the simulations was found to be key in improving the agreement with the measured cross-talk probabilities, with the other inactive materials of the modules having a smaller effect.

When cross-talk induced by single neutron-proton scattering in the first detector is kinematically possible, it is the dominant source of cross-talk.  A significant fraction of the cross-talk ($>$ 20\%) also arises from multiple $np$ and $n$C elastic scattering and becomes the dominant contribution at very large angles (close to and beyond 90$^\circ$).  At the highest energy explored here (15.5 MeV), the $^{12}\text{C}(n,n^\prime3\alpha)$ reaction becomes an additional source of cross-talk.
Such interaction mechanisms need to be taken into account in simulations and cross-talk rejection filters at these energies and beyond.

\section*{Acknowledgments}

The authors would like to thank the staff of the 4 MV VdG for providing the high-quality neutron beams.

\onecolumn

\begin{table}[htbp]
\begin{center}
\begin{tabular}{cccccc}
\hline
\hline
$E_n$ & Reaction & Beam energy & Beam intensity & Target thickness & Impregnation ratio $x$ \\
(MeV) & & (MeV) & (nA) & (mg/cm$^2$) & ($x$ = D/Ti or T/Ti) \\
\hline
\hline
1.42 & T$(p,n)^3$He & 2.37 & 500 & 0.608 & 1.59 \\
1.97 & T$(p,n)^3$He & 3.00 & 130-210& 0.956 & 1.52 \\
2.31 & T$(p,n)^3$He & 3.30 & 450-1700 & 0.608 & 1.59 \\
4.84 & D$(d,n)^3$He & 1.93 & 75-90 & 0.964 & 1.76 \\
4.88 & D$(d,n)^3$He & 1.93 & 320 & 0.929 & 1.55 \\
15.48 & T$(d,n)^4$He & 0.70 & 20-180 & 0.956 & 1.52 \\
\hline
\hline
\end{tabular}
\caption{Neutron energies used in the cross-talk measurements, reactions, and characteristics of the beams and targets employed for neutron production. Neutron energies are measured at 20$^\circ$ from the beam axis.}
\label{TabBeamsTargets}
\end{center}
\end{table}

\begin{table}[htbp]
\begin{center}
\begin{tabular}{ccccc}
\hline
\hline
$E_n$ & Detector & $d_{AB}$ & $\theta_{AB}$ & $[\theta_{min},\theta_{max}]$ \\
(MeV) & type & (cm) & ($^\circ$) & ($^\circ$) \\
\hline
\hline
1.42 & EDEN & 86 & 51 & [37, 65] \\
& & 73 & 70 & [56, 81] \\
& & 67 & 98 & [91, 103] \\
\hline
1.97 & MONSTER & 84 & 49 & [34, 63] \\
& & 87 & 67 & [54, 77] \\
& & 81 & 89 & [80, 96] \\
\hline
2.31 & EDEN & 86 & 51 & [37, 65] \\
& & 73 & 70 & [56, 81] \\
& & 67 & 98 & [91, 103] \\
\hline
4.84 & EDEN & 86 & 51 & [37, 65] \\
& & 73 & 70 & [56, 81] \\
\hline
4.88 & MONSTER & 84 & 48 & [34, 63] \\
& & 87 & 67 & [54, 77] \\
\hline
15.48 & MONSTER & 156 & 47 & [38, 57] \\
& & 153 & 65 & [57, 73] \\
\hline
\hline
\end{tabular}
\caption{Conditions of the cross-talk measurements. $E_n$: incident neutron energy on detector $A$ at 20$^\circ$; detector type; $d_{AB}$: distance between the centres of the front faces of $A$ and $B$; $\theta_{AB}$: scattering angle from the centre of the front face of $A$ to that of $B$; [$\theta_{min}$, $\theta_{max}$]: range of scattering angles covered by $B$.}
\label{Tab_CT_conditions}
\end{center}
\end{table}

\begin{table}[htbp]
\begin{center}
\begin{tabular}{ccccc}
\hline
\hline
Detector part & Detector type & Material & Composition & Density \\
 & & & & (g/cm$^3$) \\
\hline
Scintillator & EDEN, MONSTER & Xylene & C$_8$H$_{10}$ & 0.87 \\
\hline
Scintillator cell & EDEN & Stainless steel & Fe$_{74}$Cr$_{18}$Ni$_8$ & 8.0 \\
 & MONSTER & Aluminium & Al & 2.7 \\
\hline
Cell window & EDEN & Glass & O:0.460 Na:0.096 Si:0.337 Ca:0.107 & 2.4 \\
 & MONSTER & Quartz & SiO$_2$ & 2.65 \\
\hline
Light guide & EDEN, MONSTER & PMMA & C$_5$H$_8$O$_2$ & 1.19 \\
\hline
Container & EDEN & Iron & Fe & 7.87 \\
& MONSTER & Aluminium & Al & 2.7 \\
\hline
\hline
\end{tabular}
\caption{Compositions and densities of the materials included in the GEANT4 simulations. The composition of glass is given in terms of mass fractions.}
\label{Tab_Materials}
\end{center}
\end{table}

\begin{table}[htbp]
\begin{center}
\begin{tabular}{ccccccc}
\hline
\hline
$E_n$ & $d_{AB}$ & $\theta_{AB}$ & $L_{thr}$ & $P_{CT,exp}$ & $P_{CT,sim}$ (GEANT4 -- Full) & $P_{CT,exp}/P_{CT,sim}$ \\
(MeV) & (cm) & ($^\circ$) & (keVee) & [$\times 10^{-4}$] & [$\times 10^{-4}$] \\
\hline
\hline
1.42 & 86 & 51 & 75 & $2.26 \pm 0.10$ & $5.63 \pm 0.09$ & $0.401 \pm 0.019$ \\
& 73 & 70 & 75 & $0.476 \pm 0.052$ & $0.964 \pm 0.026$ & $0.492 \pm 0.055$ \\
& 67 & 98 & 75 & $0.164 \pm 0.029$ & $0.297 \pm 0.011$ & $0.553 \pm 0.097$ \\
\hline
1.97 &  84 & 49 & 100 & $4.67 \pm 0.15$ & $8.18 \pm 0.17$ & $0.571 \pm 0.021$ \\
& 87 & 67 & 100 & $0.875 \pm 0.074$ & $1.29 \pm 0.05$ & $0.679 \pm 0.062$ \\
& 81 & 89 & 100 & $0.416 \pm 0.018$ & $0.584 \pm 0.009$ & $0.711 \pm 0.033$ \\
\hline
2.31 & 86 & 51 & 100 & $5.720 \pm 0.099$ & $9.48 \pm 0.09$ & $0.603 \pm 0.012$ \\
& 73 & 70 & 100 & $1.501 \pm 0.070$ & $2.55 \pm 0.06$ & $0.588 \pm 0.030$ \\
& 67 & 98 & 100 & $0.641 \pm 0.088$ & $0.959 \pm 0.037$ & $0.668 \pm 0.095$ \\
\hline
4.84 & 86 & 51 & 100 & $12.76 \pm 0.49$ & $15.3 \pm 0.5$ & $0.834 \pm 0.041$ \\
& 73 & 70 & 100 & $5.77 \pm 0.37$ & $7.92 \pm 0.35$ & $0.728 \pm 0.056$ \\
\hline
4.88 & 84 & 48 & 100 & $13.36 \pm 0.29$ & $16.3 \pm 0.3$ & $0.820 \pm 0.023$ \\
& 87 & 67 & 100 & $4.76 \pm 0.26$ & $7.15 \pm 0.21$ & $0.665 \pm 0.040$ \\
\hline
15.48 & 156 & 47 & 100 & $3.74 \pm 0.17$ & $4.02 \pm 0.15$ & $0.930 \pm 0.054$ \\
& 153 & 65 & 100 & $2.50 \pm 0.14$ & $3.13 \pm 0.12$ & $0.798 \pm 0.054$ \\
\hline
\hline
\end{tabular}
\caption{Measured and simulated cross-talk probabilities, $P_{CT,exp}$ and $P_{CT,sim}$, respectively, and their ratio $P_{CT,exp}/P_{CT,sim}$, for each set of neutron incident energy $E_n$, relative placement of detectors $A$ and $B$ (distance $d_{AB}$ and angle $\theta_{AB}$, see Fig. \ref{Fig_detectors}), and threshold on light output $L_{thr}$. Statistical uncertainties are indicated. In the case of $P_{CT,exp}$ they include the contribution from background subtraction. The simulations, performed with GEANT4, included the scintillator, the cell walls and glass window, the light guide and the casing of each detector.}
\label{Tab_PCT}
\end{center}
\end{table}

\begin{table}[htbp]
\begin{center}
\begin{tabular}{cccccccccc}
\hline
\hline
$E_n$ & $\theta_{AB}$ & $np$ & $(n,n'3\alpha)$ & $(np)^2$ & $np$+$n$C & $(n,n'3\alpha)$+$np$ & $(np)^3$ & $(np)^2$+$n$C & $(np)^{4+}$ \\
(MeV) & (°) & (\%) & (\%) & (\%) & (\%) & (\%) & (\%) & (\%) & (\%) \\
\hline
 1.97 & 49 & $\mathbf{78}$(2) &       0 & $\mathbf{12}$(1) & 5.1(5) &      0 &  1.0(2) &  2.0(3) & 0.2(1) \\
 1.97 & 67 & $\mathbf{24}$(2) &       0 & $\mathbf{32}$(2) & $\mathbf{21}$(2) &       0 &  3.9(7) &   $\mathbf{10}$(1) & 0.7(3) \\
 1.97 & 89 & $\mathbf{22}$(1) &       0 & $\mathbf{10}$(1) & $\mathbf{33}$(1) &       0 &  3.4(3) & $\mathbf{17}$(1) & 1.2(2) \\
\hline
 4.84 & 51 & $\mathbf{68}$(3) &       0 & $\mathbf{11}$(1) &   9(1) &        0 & 1.7(4) &  3.7(6) & 0.5(2) \\
 4.84 & 70 & $\mathbf{50}$(4) &       0 & $\mathbf{14}$(2) & $\mathbf{15}$(2) &        0 & 2.7(7) &     8(1) & 1.3(5) \\
\hline
15.48 & 47 & $\mathbf{60}$(4) &  9(1) &   7(1) & $\mathbf{10}$(1) & 1.1(4) & 0.4(2) & 3.5(7) & 0.7(3) \\
15.48 & 65 & $\mathbf{49}$(3) & $\mathbf{11}$(1) &  8(1) &   9(1) & 2.0(6) & 1.6(5) & 1.9(5) & 1.0(4) \\
\hline
\hline
\end{tabular}
\caption{Fractions of cross-talk events contributed by different interaction processes in detector $A$, obtained with the GEANT4 simulations including the scintillator volume, the scintillator cell, the light guide, the window and the detector container. The fraction indicated for a multiple interaction is the sum of the various combinations of the single interactions, regardless of their order. $np$ and $n$C indicate $n+p$ and $n+C$ elastic scattering, respectively. $(np)^\nu$ indicates $\nu$ successive $n+p$ elastic scatterings. Fractions of 10 \% or higher are indicated in bold. The errors are statistical.}
\label{Tab_SimProcesses}
\end{center}
\end{table}

\end{document}